# "Secular and Rotational Light Curves of 6478 Gault"

Ignacio Ferrín
Faculty of Exact and Natural Sciences
Institute of Physics, SEAP, University of Antioquia,
Medellín, Colombia, 05001000
ignacio.ferrin@udea.edu.co

Cesar Fornari
Observatorio "Galileo Galilei", X31
Oro Verde, Argentina

Agustín Acosta
Observatorio "Costa Teguise", Z39
Lanzarote, España

Number of pages 23

Number of Figures 10

Number of Tables 6

**Abstract**

We obtained 877 images of active asteroid 6478 Gault on 41 nights from January 10$^{th}$ to June 8$^{th}$, 2019, using several telescopes. We created the phase, secular and rotational light curves of Gault, from which several physical parameters can be derived. From the phase plot we find that no phase effect was evident. This implies that an optically thick cloud of dust surrounded the nucleus hiding the surface. The secular light curve (SLC) shows several zones of activity the origin of which is speculative. From the SLC plots a robust absolute magnitude can be derived and we find $m_V(1,1,\alpha) = 16.11\pm0.05$. We also found a rotational period Prot = 3.360±0.005 h and show evidence that 6478 might be a binary. The parameters of the pair are derived. Previous works have concluded that 6478 is in a state of rotational disruption and the above rotational period supports this result. Our conclusion is that 6478 Gault is a suffocated comet getting rid of its suffocation by expelling surface dust into space using the centrifugal force. This is an evolutionary stage in the lifetime of some comets. Besides being a main belt comet (MBC) the object is classified as a dormant Methuselah Lazarus comet.

Key words: comets, minor planets, 6478 Gault

## 1.1 6478 Gault: Introduction

In the first three months of 2019 five new members have been added to the list of Active Asteroids (AA), one of them, 6478 Gault, with a curious, out of the ordinary tail. In this work we study the phase, secular and rotational light curves of this object. The concept of active asteroid has been introduced to describe asteroids that exhibit cometary activity (Jewitt et al., 2015).

The discovery of activity on this main belt asteroid was made possible by the Asteroid Terrestrial-Impact Last Alert System, ATLAS, (Smith et al., 2019). The appearance of 6478 Gault immediately called attention because the nucleus did not exhibit a coma and the tail was very long and thin with no evidence of gas. This morphology is quite different from the usual comets that exhibits a gaussian shaped coma and a gas and dust tail that expands and fades with distance to the nucleus. This tail recalls the morphology of 133P/Elst-Pizarro (Hsieh et al., 2004) with which it shares some similar features (like for example the rotational period; Table 4). Theoretical interpretations of 6478 Gault (Chandler et al., 2019; Hui et al., 2019; Jewitt et al., 2019; Kleyna et al., 2019; Moreno et al., 2019; Ye et al., 2019), agree that this object is in a state of rotational disruption. Our determination of a rotational period of 3.360±0.005 h supports this conclusion.

6478 Gault is a km-sized asteroid in the Phocaea family (Nesvorny, 2015). Table 3 gives the parameters determined in this work. Table 4 gives the rotational periods of 6478, 133P and 62412 for comparison. And Table 5 gives the orbital elements and Tisserand Parameter ( Kresak, 1982). Asteroids have $T_J > 3.0$, while comets of the Jupiter family have $2.0 < T_J < 3.0$. Clearly 6478 is an asteroid. Images of 6478 Gault exhibiting its thin, long gas-less tail are shown in Figure 1. Additional images can be found at the following link: http://www.aerith.net/comet/catalog/A06478/2020-pictures.html

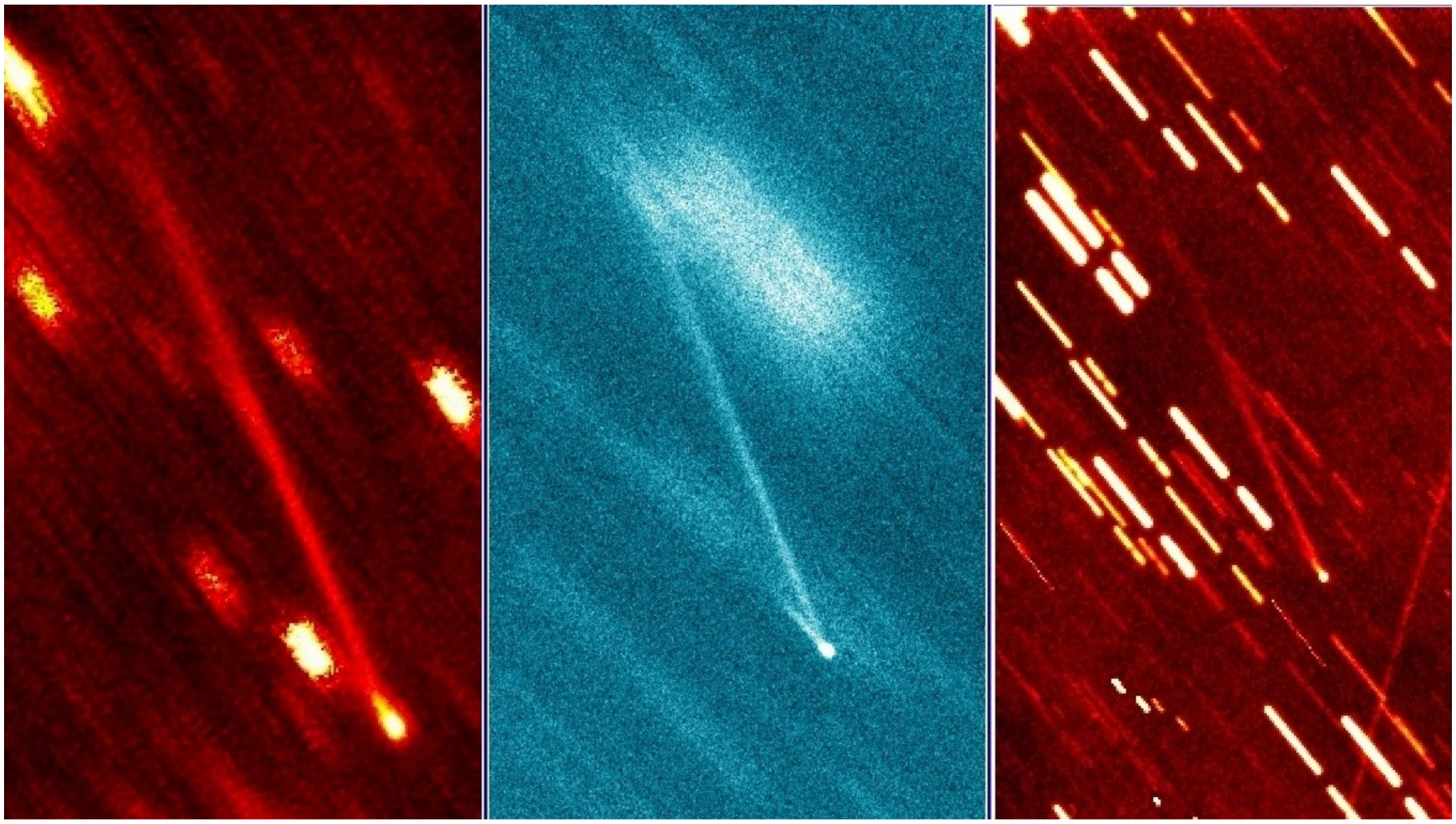

**Figure 1.** 6478 Gault. **Left image.** Median image of 15 frames x 5 min = 75 min of exposure time, V-band, taken with the 1 m f/5 reflector telescope of the National Observatory of Venezuela, on 2019 02 05. The characteristic features, star-like head, thin, long, gas-less persistent dust tail, and no tail spread as a function of distance, are all exhibited in this object. Tail B, shorter and to the left of the main tail, is compromised by a star. (IF, this work, ONV, CIDA, observatory 303). **Center image.** Image taken with a 35 cm telescope on 2019 02 13 (C. Fornari, observatory X31). 44 images of 5 minutes each were median combined for a total exposure time of 220 min . Tail B is clearly seen. On the original image the ratio length/width > 80. **Right image.** Secured on 2019 02 13 with a 24 cm telescope. 20 images x 5 min = 100 min using average combine (A. Acosta, observatory Z39). The bright patches are stars imperfectly erased by the median.

<> <> <> <> <>

**1.2 6478 Gault: Observations**

We carried out photometry of 6478 Gault on 877 images acquired on 41 nights using several telescopes listed in Table 1. The photometry log is listed in Table 2 and values are plotted in Figures 3 and 4. We are confident that our photometry is of high quality because we have adopted special procedures in that direction:

(1) Images were processed with bias, darks and flats in the usual manner.
(2) We adopt large photometric apertures typically 4-5 FWHM in order to extract the whole flux from the objects.
(3) We use from 5-15 comparison stars, and reject those that are outliers.
(4) We use APASS, the best V-magnitude catalog that spans the whole sky, created by the AAVSO to calibrate variable stars observations.

(5) We measure the observed magnitude from long series of images, typically containing 10-50 images with 3-5 minutes of exposure time each, reaching to total exposure times of hours, to plot one single photometric data point per night.
(6) Before combining, each image is examined for image artifacts, cosmic rays, clouds, star contamination, disappearance of the object, reflections and nearby bright stars. Unsuitable images are discarded before combining.
(7) The combination algorithm is the median, to clean up cosmic rays and defects.
(8) Finally, we adopt the envelope of the data points as the correct interpretation of the light curve. The envelope of the data points represents a perfect instrument + CCD, with a perfect atmosphere, capable of transmitting the whole flux from the object and thus the maximum relative brightness (Ferrín, 2005).

As a result we routinely achieve an error of ±0.01 to ±0.02 magnitudes equivalent to 1-2% error (confirmation Table 2). The error is so small that it is contained inside the plotting symbol in all of our plots.

We also used the MPC observations data base. This database is of lower quality because its photometry is a byproduct of astrometry and several non-photometric catalogs and filters are used (Ferrín, 2010). To decrease the scatter and increase the signal to noise ratio, we apply a 5 data point running mean. These values may have to be moved up or down a little bit to fit the high precision photometry of our work.

**1.3 6478 Gault: Literature search**

There have been 6 articles up to the moment of this writing, related to 6478 Gault. What follows is a resume of some numbers of interest to this investigation, not a review of these complex and sophisticated papers.

(1) Ye et al. (2019) describe two brightening events that start on 2018 October 18±5 d, and a second one that started on 2018 December 24±1 d, and that released 2x10^7 kg and 1x10^6 kg respectively. Each event persisted for about a month. Dust dynamics showed grains of up to 10 microns in size ejected at velocities less than 1 m/s regardless of particle size. Additionally they derive an upper limit to the ejection velocity of < 8 m/s. These ejection events are clearly defined in our Secular Light Curve (Figures 3 and 4, active zones Z1 and Z2).

(2) Kleina et al. (2019) determined an absolute magnitude of 14.4 in the V band based on ~1000 survey observations, and with an adopted albedo of 0.04, representative of comets, derive a nucleus 9 km in diameter. They also adopt a density of 3000 kg/m$^3$. Although they have an extensive series of observations these did not show a light curve and the rotational period could only be estimated at ~1h for one peak, and ~2h for two peaks. They derived a $v_{eject}$ = 0.7 m/s for the maximum emission velocity.

(3) Moreno et al. (2019) adopt for the dust density 3400 kg/m$^3$ which is appropriated for S-type asteroids. They also adopt a geometric albedo $p_V$ =0.15 typical of S-types. With an adopted absolute magnitude of $H_V$ = 14.4, they find a diameter of 4.5 km. They also apply a phase coefficient of 0.033 mag/°. Long series of observations secured with the TRAPPIST-North and South telescopes did no exhibit evidence of a rotational signature.

(4) Jewitt et al. (2019). These authors find mass loss rates of ~10-20 kg/s and a typical particle radius in the main tail <a> ~ 1000 μm. Using the thickness of the tail at the time of orbital plane crossing they are able to calculate a value for the ejection velocity $v_{eject}$ ~ 2 m/s. They also find a close similarity of 6478 to comet 311P/2013 P5 who exhibited multiple dust ejections (Jewitt et al., 2018). The absence of gas was confirmed spectroscopically.

(5) Hui et al. (2019). They confirm the slow dust ejection velocity for the dust, finding $v_{eject}$ = 0.15 m/s. As previous authors they infer that the mass-loss events were caused by rotational instability. They conclude that the two tails observed were caused by the two outburst that were identified in the light curve.

(6) Chandler et al. (2019). They used archival observation to demonstrate that Gault has a long history of previous activity. Their data suggests that activity is caused by a body spinning near the rotational limit. They recognize 6478 as a new class of object perpetually active due to rotational spin up. Their data is plotted in our Figure 3.

It is important to point out that none of the six manuscripts published on this object had a value for the rotational period, thus their conclusions on the reason of activity is entirely theoretical. However this work will validate these conclusions (see Section 2.3). Also, our secular light curve (SLC) of 6478 will show that the activity is not perpetual but episodic. The reasons for being episodic are speculative at the present time.

Additional works might be relevant to this investigation.

(7) The conclusion by Hshie et al. (2004) that *"apparent low ejection velocities of optical dominant dust particles, as implied by the lack of an observable sunward dust extension or coma-like dust halo around the nucleus, and the narrow width of the dust trail, make extremely small particles unlikely to be optically significant"*, in agreement with Jewitt et al. (2019) who also found large particles. These results are consistent with the hypothesis that these large particles are the remnant of the cometary activity deposited on the surface. It is also consistent with the idea that these objects might be completely covered with a thick mantle or dust layer.

(8) Kokotanekova et al. (2017) studied an ensemble of Jupiter-family comets, finding a cut-off in bulk densities at 0.6 g/cm$^3$ if the objects are strength less. They also find an increasing linear phase function coefficient with increasing albedo. Their median linear phase function coefficient for JFCs was 0.046 mag/deg and their median geometric albedo was 4.2%. These values are useful when this information is not available for a given object. We will use their albedo as a lower limit in our calculations.

(9) Hsieh et al. (2009) note that low albedo ($p_R$ < 0.075) remains a consistent feature of all cometary (i.e., icy) bodies, whether they originate in the inner solar system (the main belt comets, MBCs) or in the outer solar system (all other comets). We will use their albedo as an upper limit in our calculations.

**2.1 6478 Gault: Phase and SLC plots**

To advance in our understanding of this object we will make use of the phase plot and the secular light curve (SLCs, Ferrín, 2010a, 2010b). The phase plot shown in Figure 2, depicts the absolute magnitude uncorrected for phase vs phase angle, and does not exhibit a dependence on phase, implying that the surface is not seen due to an optically thick layer of dust surrounding the nucleus. Accordingly no phase correction has been applied to our data. In those cases in which the phase effect is not detected, the absolute magnitude has to be defined using only the SLC plot. The SLC formalism depicts the absolute magnitude of the object versus time, from aphelion –Q, to aphelion +Q (Figures 3 and 4). The SLC of 6478 shows evidence of 6 zones of activity, Z1 to Z6. Z1 and Z2 are the outburst studied by (Chandler et al., 2019; Hui et al., 2019; Jewitt et al., 2019; Kleyna et al., 2019; Moreno et al., 2019; Ye et al., 2019).

The origin of these active zones is controversial. It might looks like the orbital plane crossings (PCs) are related to the peaks, but this interpretation is not completely convincing. A plane crossing produces an enhancement of the tail lasting typically a few days. However the six zones of activity are much wider than a few days, putting in doubt this interpretation. Active zones Z1 and Z2 may have been the result of an impact because their amplitude is much larger than that of zones Z3-Z6 and their duration is shorter (confirmation Table 3).

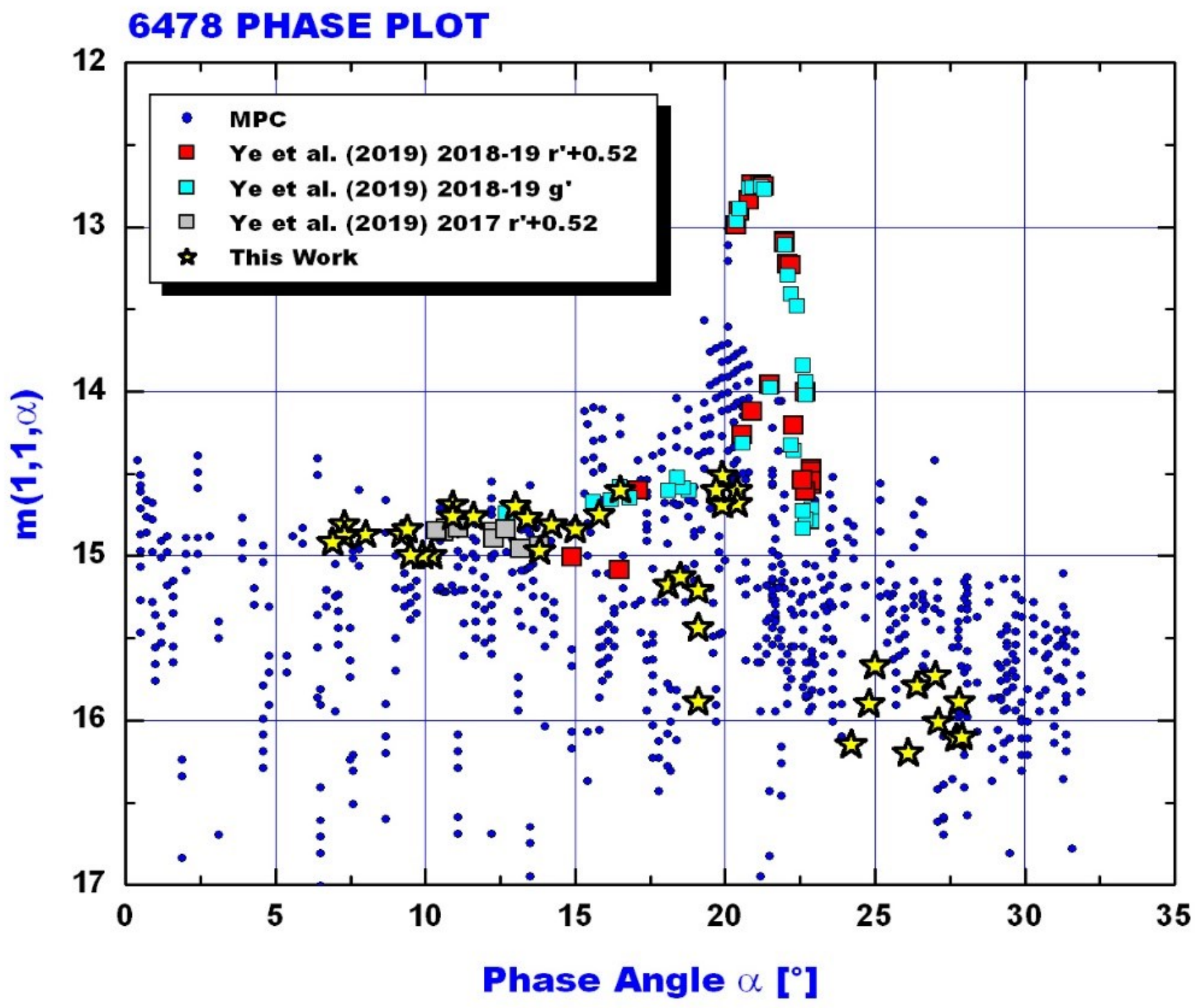

**Figure 2** Phase plot of 6478 Gault. The comet does not exhibit a phase effect, implying that we are not seeing the surface. The object must be surrounded with an optically thick cloud of dust that hides the nucleus. That is the reason why we have not applied any phase correction to our observational data. Most of the data shows the comet in an excited state, thus to define the absolute magnitude we have to use the SLC plot.

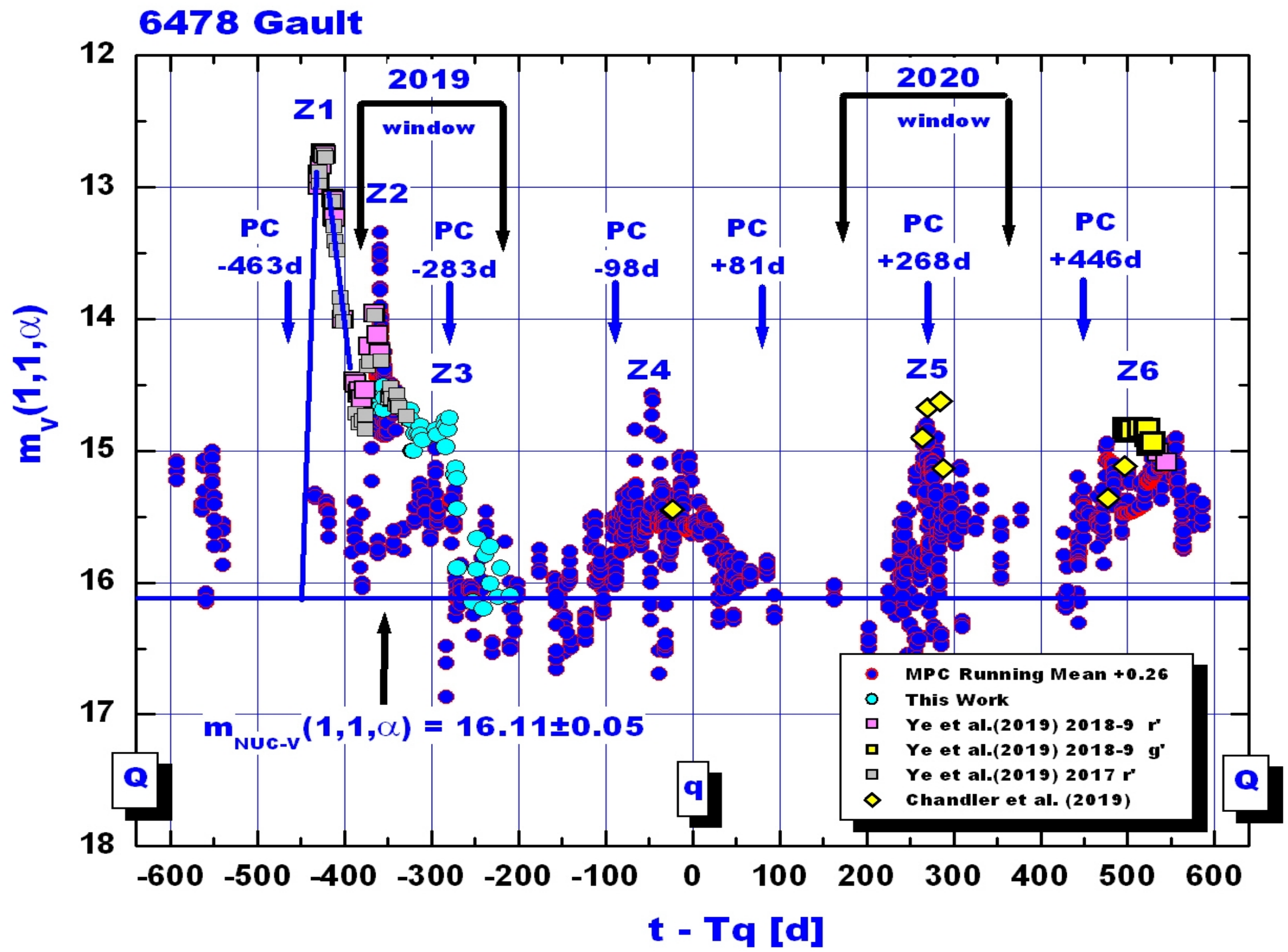

**Figure 3.** SLC of 6478 Gault. To smooth out the scattering of the data and to increase the signal/noise ratio we applied to the MPC database a 5 data points running mean. Different data sets were combined to create this plot. The absolute magnitudes listed by the Minor Planet Center and the JPL Small Body Database Browser, $H_V$= 14.3 and $H_V$= 14.4 respectively, are too bright. A new value $m_V(1,1,\alpha)$ = 16.11 ±0.05 is derived in this work using the mean value of our five faintest measurements. Notice that this result is consistent with the faint values at t-Tq = +160 d. The object shows 6 zones of activity, Z1 to Z6, the reason for which is not clear at the present time. PC = Orbit Plane Crossing, indicate when the Earth crossed the orbital plane. The plane crossings do not seem to be correlated with the activity zones. Above, the observing windows in 2019 and 2020 are shown. The 2020 window will allow observations of the Z5 zone. Notice that the observations compiled by Chandler et al. (2019) agree quite well with active zones Z4, Z5 and Z6.

<> <> <> <> <>

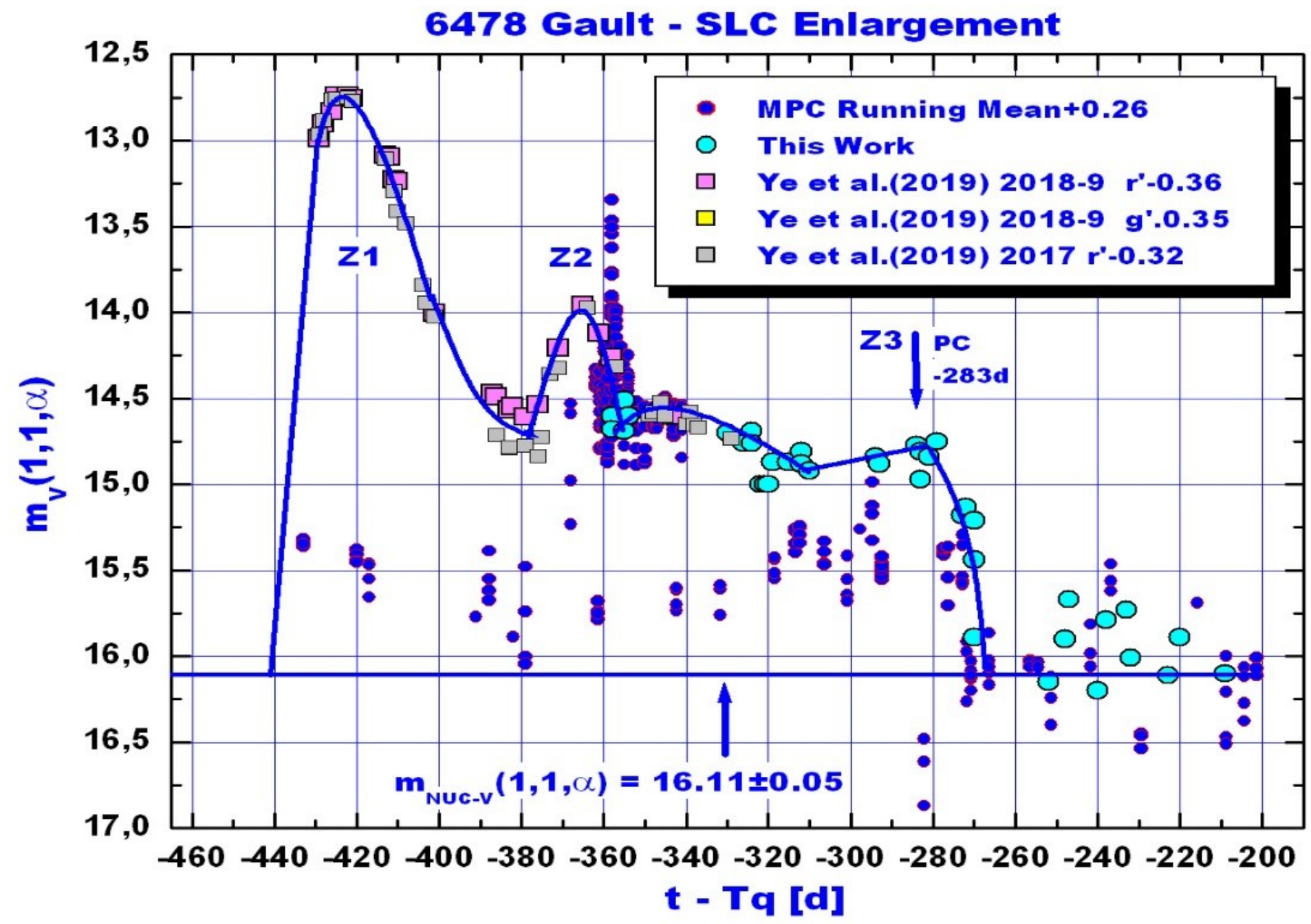

**Figure 4.** SLC of 6478 Gault. Enlargement of the region -465 d to -200 d before perihelion. The two outburst Z1 and Z2 can be clearly seen and delimited. The average of the 5 faintest measurement gives our best estimate of the absolute magnitude of the nucleus.

<> <> <> <> <> <>

## 2.2 6478 Gault: Absolute Magnitude, Diameter, Mass

**Absolute magnitude.** The SLC plot gives the whole picture from –Q to +Q (Q = aphelion distance) needed to select a reliable absolute magnitude. The $m_V(1,1,0)$ and the observed magnitude $m_V(\Delta,R,\alpha)$ are related by

$$m_V(1,1,0) = m_V(\Delta,R,\alpha) - 5.\text{Log}\,(\Delta.R) - \beta.\,\alpha \qquad (1)$$

where Δ is the object-Earth distance, R the Sun-object distance, both in astronomical units, α is phase angle in degrees, and β the phase coefficient.

In Figures 3 and 4, the SLC plot, five data points have the lowest values of the dataset, 16.15, 16.20, 16.01, 16.10, and 16.11. The range is 0.19 in excellent agreement with the amplitude of the rotational light curve found later on (Amp=0.16±0.02) (see Section 5.3). The mean value is our best estimate of the absolute magnitude, $m_V(1,1,\alpha)$ = 16.11±0.05.

The works cited in Section 1.3 adopted for their calculations an absolute magnitude 1.71 magnitudes brighter than our result. Consequently some of the model calculations performed in the above works might have to be revised.

**Diameter.** With this information it is possible to calculate the diameter D if the geometric albedo, pv, is assumed. Jewitt (1991) gives the following formula

$$p_V .r_{NUC}^2 = 2.24x10^{\wedge}22.R^2.\Delta^2\ 10^{\wedge}[0.4(m_{SUN} - m_V(1,1,0))] \quad (2)$$

which can be written in a more amicable form

$$Log\ [\ p_V.D^2/4\ ] = 5.654 - 0.4\ m_V(1,1,0) \quad (3)$$

Kokotanekova et al. (2017) calculated a mean geometric albedo of the Jupiter family of comets $<p_V> = 0.042\pm0.005$. Since the SLC of 6478 indicates that the object is behaving more like a comet than an asteroid, we are going to adopt this value in our calculations. Then we find for the diameter D = 3.93±0.25 km, r = 1965±125 m. Hisieh (2009) gives a maximum $p_V = 0.075$. If we use this value we find D = 2.94 km, r = 1470 m. This calculation assumes that the object is single. However in Section 2.4 we will show that it is a binary.

**Mass.** To calculate the mass we need a density. Later on we will find a rotational period for Gault Prot = 3.360±0.005 h and an amplitude Amp= 0.16±0.02, which translates to a ratio of axis a/b = 1.16. In the Section on rotation (Section 2.3) we calculate that the minimum density to hold material at the tip of this rotating asteroid is $P_{CRITICAL} \sim 1120$ kg/m$^3$ (Jewitt et al., 2014). Then the estimated mass of the nucleus is $M_{NUC}$ = (3.6±0.35) x 10^13 kg for $p_V = 0.042$ or $M_{NUC}$ = (1.5±0.25) x 10^13 kg for $p_V = 0.075$.

The masses ejected calculated by the previous researchers were 2 to 4x10^7 and 1x10^6 kg for active zones Z1 and Z2 (Ye et al., 2019; Jewitt et al., 2019). If this mass were to be released by apparition and sustained in time, the number of returns left would be ~18.000. If this mass were spread out uniformly over the surface, the depth of the layer would be only 0.4 mm. This information combined with the negative detection of a gas implies that there is a deep layer of dust on this surface which is quenching the sublimation activity. Thus this object is actually dormant.

## 2.3 6478 Gault: Rotational Period

An attempt to measure the rotational period was made by Kleyna et al. (2019) who found a value ~2h. However phasing and smoothing of their data did not reveal any obvious light curve, suggesting that the periodic signal was buried in aperiodic, non-Gaussian noise.

We used the 24 cm telescope of Z39 to take 178 observations of asteroid-comet 6478 Gault and processed them with software Canopus and Maxim-DL to measure the rotational period of this object (Ferrín and Acosta, 2019) (the data is listed in Table 6). We observed 6478 from January 12th to April 8th during 8 nights and used a 5 data point running mean to diminish the scattering of the data and to increase the signal to noise ratio by a factor of 2.2. Since this average operates only in the vertical direction there is no way it can modify the period. The NASA Exoplanet periodogram tool was used to calculate the period with the Lomb-Scargle option. To our surprise the periodogram found a period around ~0.07 d corresponding to rotational light curve with one peak and Prot(1)=1.680±0.002 h and amplitude Amp(1) = 0.18±0.02 mag. A two-peaked light curve would have Prot(2)=3.360±0.005 h and amplitude Amp(2) = 0.16±0.02. The periodogram is shown in Figure 5. The first period with one rotational peak violates the rotational limit for disruption of a rubble pile asteroid ( ~2.2 h) (Jewitt et al., 2014), so we adopt the two-peaked light curve.

It is interesting to point out that we did not expect that the 24 cm telescope could find a period for this 17-magnitude target, but it did. There is no way in which this result could have been falsified. Each observation corresponds to a phase in the phase curve, and the observer does not have any way to know the phase for any observing night. As a confirmation, the light curve produced looks quite robust (Figures 6 and 7) and has been confirmed (Figure 9).

A 0.16 mag amplitude corresponds to a ratio of axis a/b=1.16. The minimum density needed to ensure the material at the tip of a prolate body in rotation about its minor axis that is gravitationally bound is given by Jewitt et al. (2014)

$$\rho_{CRITICAL} \sim 1000\ (3.3/Prot)^2\ [a/b] \qquad (4)$$

If we use the observed period, 3.360 h and the observed ratio a/b=1.16 then the minimum density comes out to be $\rho_{critical} > \sim 1.12$ gm/cm$^3$, a very reasonable value expected for a comet mainly made of water ice.

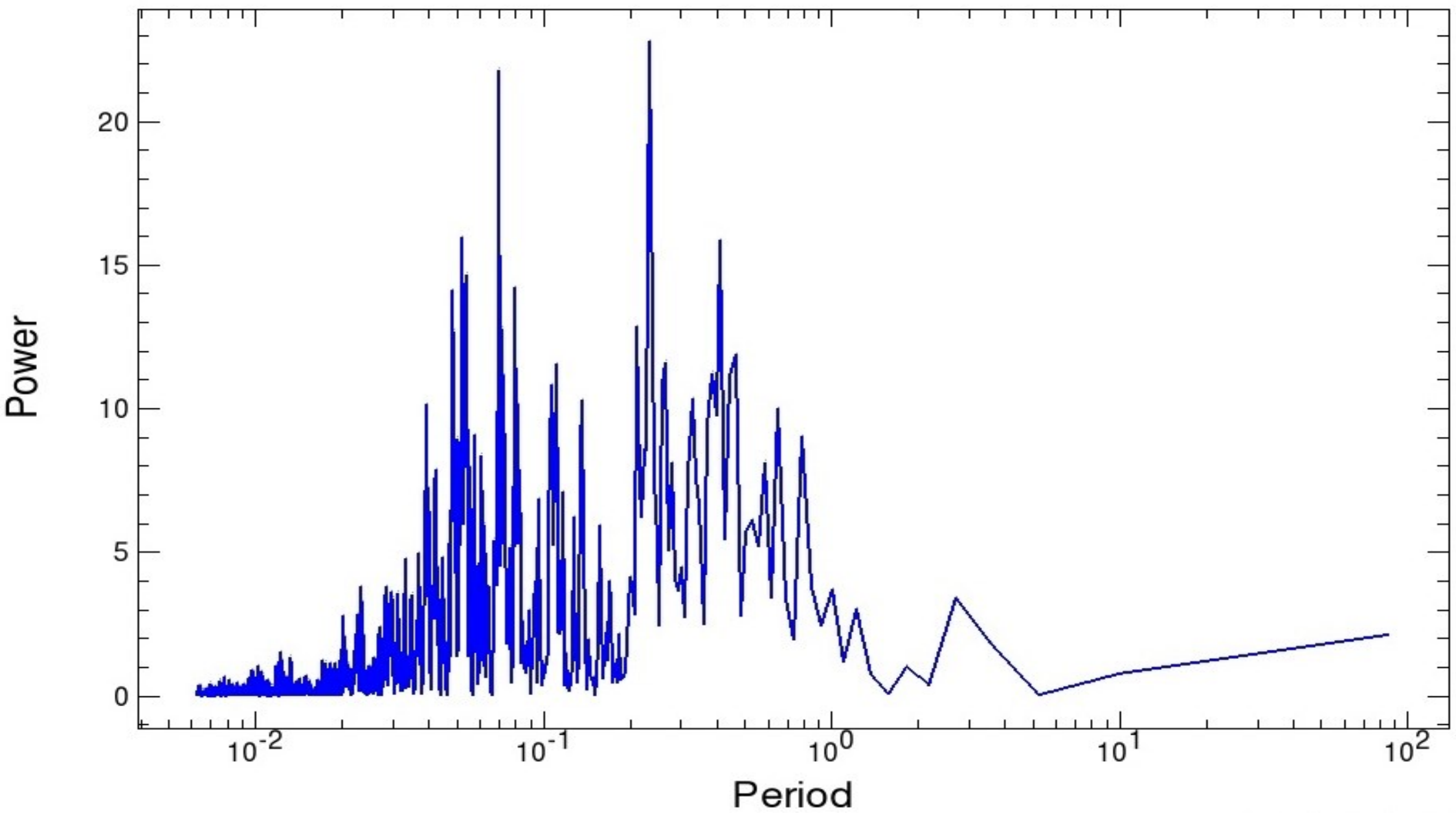

**Figure 5.** The periodogram of 6478 Gault shows two main peaks at periods ~0.07 d and ~0.23 days. The peak at ~0.23 = 5.52 h gives a light curve of poor quality so it is not considered further. Peak 1 with one rotational peak corresponds to Prot(1)=1.680±0.002 hours and with two peaks corresponds to Prot(2)= 3.360±0.005 hours. The amplitudes are Amp(1) = 0.18±0.02 and Amp (2) = 0.16±0.02 magnitudes respectively. The first one with one rotational peak violates the rotational limit for disruption of a rubble pile, so we adopt a two-peaked light curve. This plot was generated by the periodogram tool of the NASA Exoplanet page.

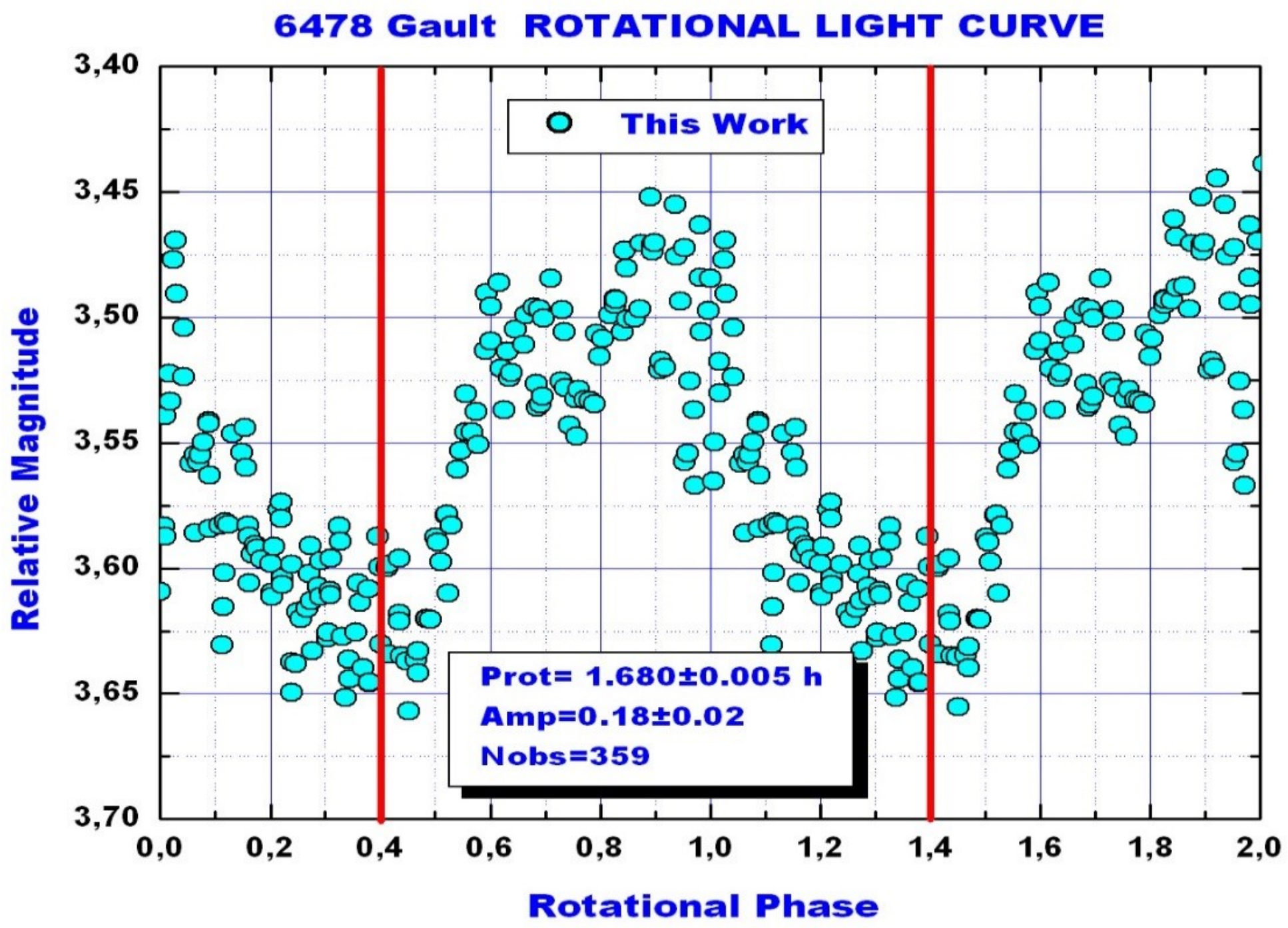

**Figure 6.** The first solution for the rotational period of 6478 Gault uses one single maximum with period Prot(1)=1.680±0.005 hours and amplitude Amp = 0.18±0.02 magnitudes but violates the rotational limit for disruption of a rubble pile (~2.2 hours).

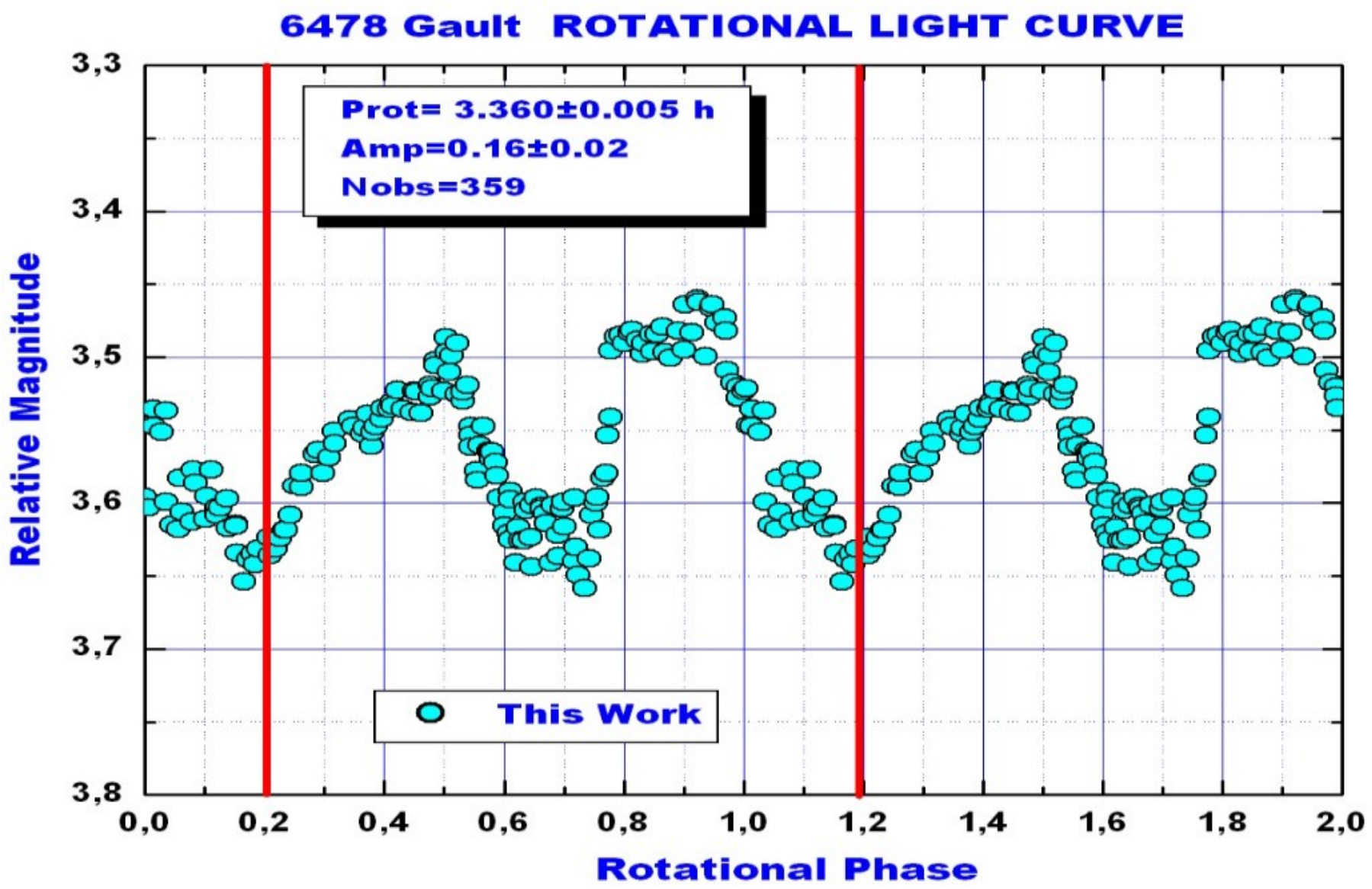

**Figure 7.** The second solution for the rotational period of 6478 Gault shows two rotational peaks and a rotational period Prot(2) = 3.360±0.005 hours with amplitude Amp(2) = 0.16±0.02 magnitudes.

In Table 4 we list the rotational periods of Gault, 133P/Elst-Pizarro and 62412. We see that all of them have almost identical values. Since they exhibit the same tail morphology of 6478, this implies that these objects are also in a state of rotational disruption.

**2.4 6478 Gault: Eclipse**

Surprisingly the phased light curve diagram shows evidence of an eclipse in the form of a flat bottom V (Figure 7), with the following de-trended values T1(phase,mag) = (0.53,0.0) = 190 deg, T2 = (0.62,0.13) = 223 deg, T3 = (0.74,0.13) = 266 deg, T4 = (0.80,0.0) = 288 deg, T4-T1 = 0.27 =97 deg, T3-T2 = 0.12 = 43 deg, (T1+T2+T3+T4)/4 = 0.673, m3-m1 = 0.13 mag (Figure 8).

To produce a flat bottom the two components have to have different sizes with radii of the primary and secondary, $r_P$ and $r_S$. An asteroid that has been evolved to a top-like shape by radiative forces due to the YORP effect has been designated as a YORPoid (Deller, 2015). 101955 Bennu, 162173 Ryugu and 2867 Steins may be good examples of YORPoids. However at the end of Section 3 we show that YORP could not have been a significant force on 6478 and thus the shape is unknown. Since the shape does not affect the result of this calculation, for a first approximation we will assume spheres and equal albedos for the pair. Then the area projected by both components

$$A_{Max} = \pi \,.\, ( r_P^2 + r_S^2 ) \tag{5}$$

$$A_{min} = \pi \,.\, r_P^2 \tag{6}$$

The duration of the eclipse is T4 – T1 = 0.907 h and it must correspond to a distance v . ( T4 – T1 ) = 2 . ($r_P + r_S$ ), thus we can calculate the projected space velocity, v. Kepler's third law agglutinates a number of measurable parameters and assuming a circular orbit of radius a around the primary (Jewitt et al., 2018):

$$a = 4.\pi.G.\rho.( r_P^2 + r_S^2 ) \;/\; ( 3.v^2 ) \tag{7}$$

An analysis depends on the absolute magnitude value of the object and requires selecting two free parameters, the geometric albedo, pv, and the density, ρ. For the geometric albedo we are going to adopt the limits set by Kokotanekova et al. (2017) (~0.042) and Hisieh et al. (2009) (<0.075). For the density we are going to adopt the minimum critical value derived from the rotational parameter (1120 kg/m$^3$). The solution for a contact binary requires an additional condition,

$$( r_P + r_S ) = 2.a \tag{8}$$

Solving equations (5), (6), (7) and (8) with $p_V$ = 0.042, 0.06, 0.075, we find a radius of the primary 1850 m, 1550 m, 1390 m, and for the secondary 660 m, 552 m and 495 m. In order to get these results a minimum density of 1590 kg/m$^3$ has to be adopted. If the density increases above this value the pair is detached.

The observational orbital velocities are $v_{orb}$ = 1.54, 1.28, 1.15 m/s in very good agreement with the theoretical scape velocity of the dust 0.7 – 8 m/s (Chandler et al., 2019; Hui et al., 2019; Jewitt et al., 2019; Kleyna et al., 2019; Moreno et al., 2019; Ye et al., 2019).

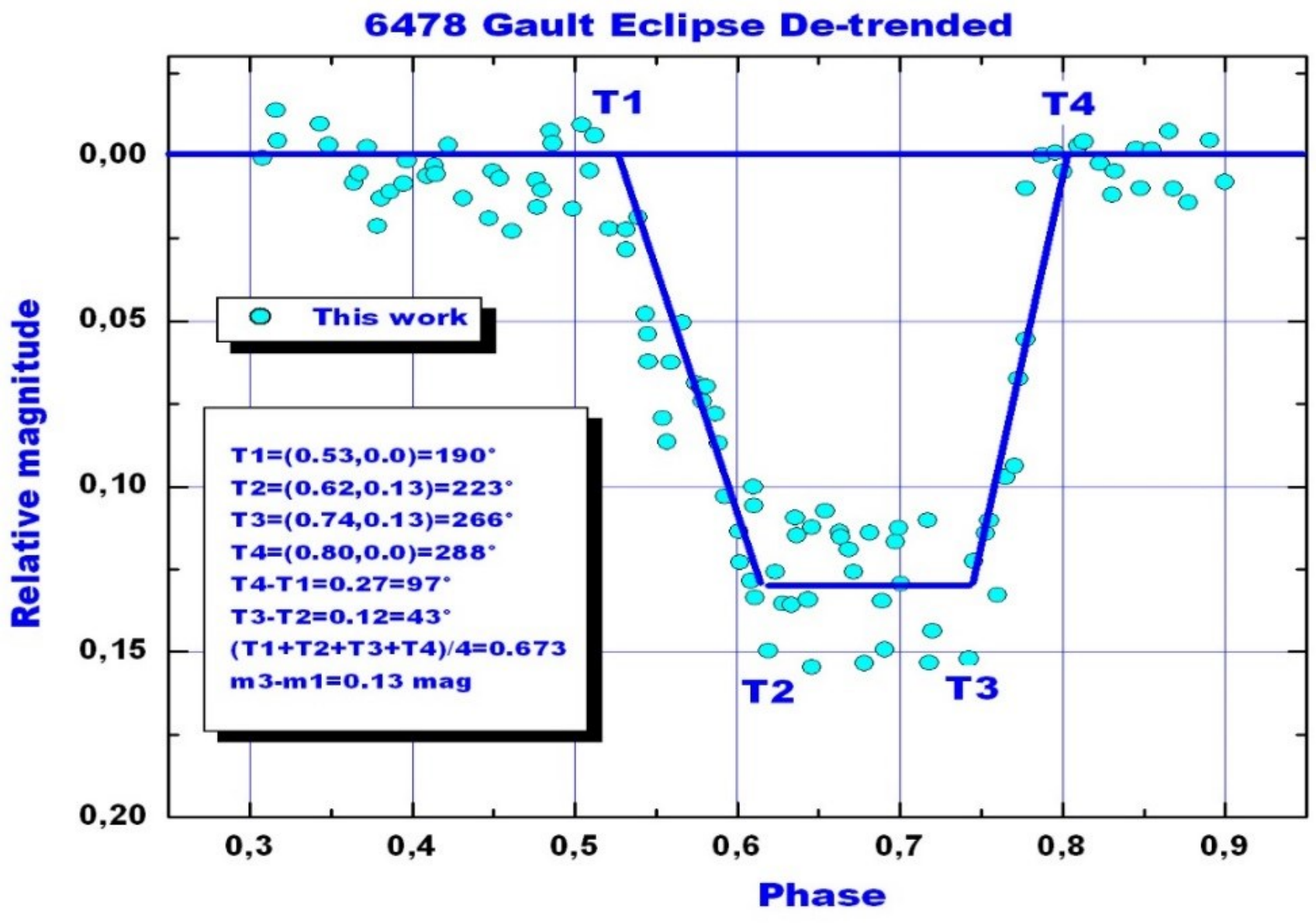

**Figure 8.** De-trended Eclipse in the light curve of 6478 Gault. The four eclipse parameters, T1, T2, T3 and T4 are listed in the inset table. The total eclipse lasted for (T4-T1)*Prot = 0.907 hours. The deduced physical parameters of the double nucleus are calculated in the text.

<> <> <> <> <>

## 2.5 6478 Gault: Confirmation of the rotational period

Carbognani and Buzzoni (2019) were able to reproduce our result (Figure 9). They used the 80 cm Telescope of the Osservatorio Astronomico della Regione Autonoma della Valle d’Aosta, Associato al INAF (Observatorio Astronomico di Torino, Italy). We applied a 5 data point running mean to their data to increase the signal to noise ratio. They found a period of 3.358 h in excellent agreement with our result. However the eclipse is no longer evident, leaving the object with one single minimum per rotation interpreted as an albedo feature.

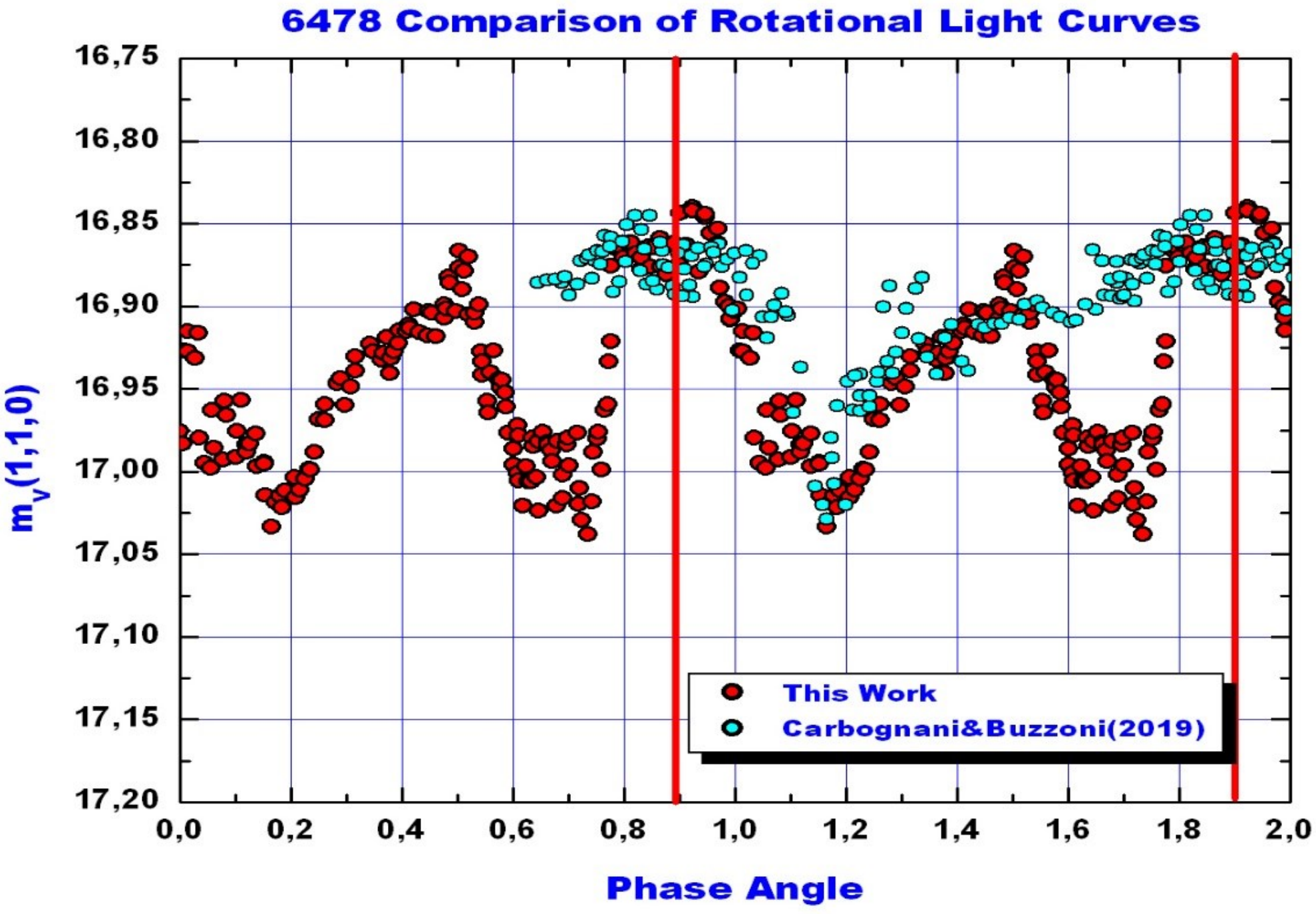

**Figure 9.** The rotational light curve of Carbognani and Buzzoni (2019) is compared with our rotational light curve. They found a period of 3.358 h vs 3.360 hour for our work. The agreement is excellent. The principal distinction is that the eclipse feature has disappeared. Then the light curve can not be that of an elongated body because it would exhibit two peaks. The new interpretation is of an almost spherical nucleus with a zone of lower albedo, producing one single minimum per revolution.

<> <> <> <> <>

## 3. 6478 Gault: Discussion and Current Understanding

From the previous works (Chandler et al., 2019; Hui et al., 2019; Jewitt et al., 2019; Kleyna et al., 2019; Moreno et al., 2019; Ye et al., 2019), all of which agree on a basic understanding of this object, and with the results presented in this investigation, the following picture emerges, and this picture may be shared with two other objects, 133P/Elst-Pizarro (Hsieh et al., 2004) and 62412 (Sheppard and Trujillo, 2014) (Tables 4 and 5).

Comets expel gas and dust from their surfaces during their lifetimes. Fine dust is carried away by the gas, but large dust falls down onto the surface creating a dust mantle that increases in thickness with time. At the same time that this process is taking place, the majority of cometary nucleus are increasing their rotational frequency due to offset gas jets that exert a torque on the nucleus diminishing its rotational period and increasing its angular momentum. After many returns around the Sun, the dust mantle increases in thickness and there comes a moment when the mantle quenches the cometary activity creating a dormant comet (Snodgrass et al., 2017).

The comet will remain in this stage if the perihelion distance is kept the same or if it increases in distance to the Sun. But the dormancy can be broken if the perihelion distance decreases, creating a thermal wave that will penetrate deeper into the nucleus reactivating the activity. The comets with decreasing perihelion distance and increasing activity have been designed as Lazarus comets (Ferrín et al., 2013). A good example of a Lazarus comet is 107P/Wilson-Harrington.

Among these old comets there may be a small group that due to the spin up may have rotational periods near the rotational limit (Hsieh et al., 2014; Jewitt et al., 2018). These objects would emit dust into space, expelled at very small ejection velocities by centrifugal forces, but would not show evidence of gas. It is believed that this is the mechanism that expelled the dust off the surface of these objects.

Due to the thick layer of dust accumulated on the surface, the activity is quenched or suffocated. By expelling dust into space, the dormant comet is actually getting rid of the suffocation. Thus our conclusion is that 6478 Gault is a dormant suffocated comet getting rid of its suffocation by expelling surface dust into space using centrifugal forces. This is an evolutionary stage in the lifetime of some comets (Ferrín et al., 2019b).

Some authors propose a YORP mechanism to explain the fast rotational period of 6478 (Chandler et al., 2019; Hui et al., 2019; Jewitt et al., 2019; Kleyna et al., 2019; Ye et al., 2019). However Samarashinha and Mueller (2018) show beyond the shadow of a doubt that rotational changes can take place in one single orbit, and in their Table 1 list a rotational period change of -21 minutes for 67P/Churyumov-Gerasimenko and +150 minutes per orbit for 103P/Hartley 2, implying that the time-scale for a rotational change due to sublimating torques is of the order of ~5 years, while YORP has a time-scale of 100 million years for a 4 km diameter comet. And Chandler et al. (2019) has shown that 6478 Gault was active in the previous return. Thus the possibility of YORP being significant for 6478, has to be ruled out.

**4. 6478 Gault: Age**

It would be interesting if we could assign an age to this comet. We can define a proxy for age based on the amount of mass lost per apparition (Ferrín, 2014). The argument is that as they age the amount of mass expelled diminishes with time. Thus a proxy for age may be the inverse of the apparition mass loss. The proposed definition is

$$\text{ML-Age [cy]} = 3.58\text{x}10^{11} \text{ kg / Mass Loss per Apparition} \quad (9)$$

where cy stands for comet years, different from Earth's years, and the constant was chosen so that comet 28P/Neujmin 1 has an age of 100 cy. For the two initial outbursts Ye et al. (2019) calculate masses released 2.7x10^7 kg and 1x10^6 kg, while Moreno et al. (2019) find 1.4x10^7 and 1.6x10^6 kg. For the first outburst Hui et al. (2019) find 9x10^6 kg. If we first take the average of these values and then scale them to the whole apparition we estimate an apparition mass loss of 2x10^7 kg. Applying Equation (9) we find ML-Age(Gault) ~ 18000 cy. Any object above 100 cy is a Methuselah comet (Ferrín, 2014), thus this is a very evolved object.

**5. Future Evolution**

The activity of 6478 Gault depends on the total amount of energy received from the Sun. In Figure 10 we show the solar insolation due to perihelion distance changes from 1900 to 2100. The energy received depends on the inverse of the perihelion distance squared, normalized to the year 2020 ( = 1 = 100% ). We see that the perihelion distance has a secular decrease, causing a secular increase in energy input from the Sun, and classifying this object as a Lazarus comet. The conclusion is that the current activity will remain quite stable, increasing secularly at a rate of only +0.002 % per year.

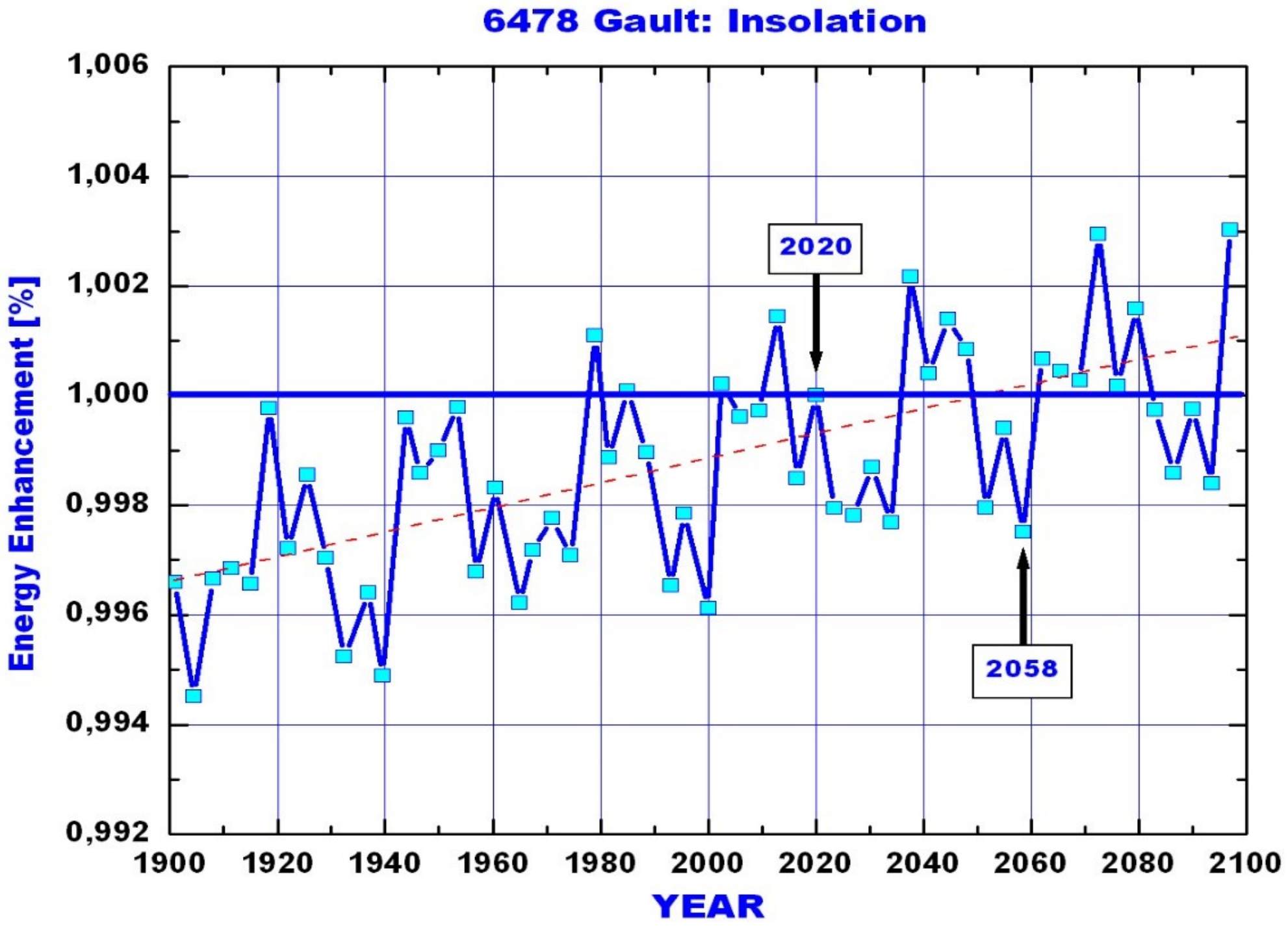

**Figure 10.** 6478 Gault: insolation . The small vertical variation of the energy received from the Sun indicates that the orbit and energy budget of this object is quite stable in the time range covered by this plot (1900-2100). The decrease from 2020 to 2058 (-0.022%) might be sufficient to quench the activity. The tilted line indicates a secular increase in solar insolation due to a diminution of perihelion distance, classifying this object as a Lazarus comet.

<> <> <> <> <>

**6. Conclusions**

(1) We present the phase curve of 6478 Gault and find that there is no phase effect. We interpreted this as due to an optically thick cloud of dust surrounding the nucleus and hiding the surface.

(2) The SLC of this object exhibits the existence of six zones of activity which complicate the understanding of its secular activity. An explanation based on the orbital plane crossing is not entirely satisfactory, thus the reason for 6 active zones is speculative. Zones 1 and 2 might have been due to an impact.

(3) We determine the absolute magnitude of 6478 as $m_V(1,1,\alpha)$= 16.11±0.05.

(4) We determined the rotational period of the object. The periodogram shows one period Prot(1)=1.680±0.002 hour with one single rotational peak of activity that violates the rotational limit for disruption of a rubble pile (~2.2 hours) and is rejected. Doubling that period we get Prot(2)= 3.360±0.005 hours with a two-peaked light curve, similar to the rotational periods of comets 133P/Elst-Pizarro (3.471 h) and 62412 (3.33 h) to which it resembles morphologically. The Amp(2) = 0.16±0.02 magnitudes.

(5) The rotational light curve shows evidence of an eclipse. We find for the radius of the primary 1850 m, 1550 m, 1390 m, and for the secondary 660 m, 552 m and 495 m for geometric albedos 0.042, 0.06, 0.075 repectively. In order to get these results a minimum density of 1590 $kg/m^3$ has to be adopted. If the density increases above this value the pair is detached.

(6) These objects are in a state of flux. Their changing perihelion distance modifies the amount of energy received from the Sun. In the case of 6478 the perihelion distance is decreasing secularly and thus the insolation is increasing although at a very slow pace. Thus the comet is rejuvenating and classified as a Lazarus comet.

(7) Our rotational period for 6478 Gault was confirmed by Carbognani and Buzzoni (2019) who found Prot = 3.358±0.005 hours in excellent agreement with our result. However in their light curve the eclipse is no longer visible, leaving the light curve with one single minimum per rotation. This suggests a spherical shape with one single dark region, an albedo feature.

(8) Using a proxy for age, we calculate that this object has a ML-Age ~ 18000 cy (comet years). Any object above 100 cy is a Methuselah comet (Ferrín, 2014), thus this is a very evolved object.

(9) Our conclusion is that 6478 Gault is a dormant suffocated comet getting rid of its suffocation by expelling surface dust into space using centrifugal forces. This is an evolutionary stage in the lifetime of some comets.

(10) Besides being a MBC this object is classified as a dormant Methuselah Lazarus comet.

## 7. Acknowledgements

We want to thank an anonymous referee for his many suggestions that improved the scientific value of this manuscript significantly. This work contains observations made at the National Observatory of Venezuela, ONV, Centro de Investigaciones de Astronomía, CIDA, Mérida. The help of Giuliatna Navas and the night assistants Richard Rojas, Leandro Araque, Dalbare Gonzalez, Fredy Moreno, Carlos Pérez, Gregory Rojas, Daniel Cardozo and Ubaldo Sánchez, is particularly appreciated. We used as comparison stars the Photometric All Sky Survey (APASS) catalogue published by the American Association of Variable Stars Observers (AAVSO), funded by the Robert Martin Ayers Sciences Fund and NSF AST-1412587. We acknowledge with thanks the comet observations from the COBS Comet Observation Database, contributed by observers worldwide and used in this research. The contribution of images by Jesús Canive, Edwin Quintero and José Francisco Hernández is appreciated.

# TABLES

**Table 1.** Instrumentation

| Observatory | MPC Code | City | Country | Diameter | CCD | Observer |
|---|---|---|---|---|---|---|
| Galileo Galilei<br>Lat-31°49'22.96"<br>Lon60° 31' 14.3" | X31 | Oro Verde Entre Rios | Argentina | 35 cm | Sbig 8300M | Cesar Fornari |
| Costa Teguise<br>Lat +28° 59 45.8<br>Lon 13° 30' 04.7 W<br>Alt 37 m | Z39 | Lanzarote, Canary Islands | Spain | 24 cm | ST-8XME | Agustin Acosta |
| CIDA<br>Lat 8° 47'<br>Lon 70° 52' W<br>Alt 3590 m | 303 | Merida | Venezuela | 100 cm | Techtronic | Ignacio Ferrín |
| Ibiernes<br>Lat 41° 05' 51"<br>Lon -02° 34' 22"<br>Alt 1058 m | ------ | Guijosa | Spain | 30 cm | RC 12" F8 SBIG STL 1301 3 CCD | Jesus Canive |
| Altamira<br>Lat +28° 14' 46"<br>Lon -16° 26' 32"<br>Alt 589 m | X05 | Fasnia, Canary Islands | Spain | 40 cm | 1336x890 SBIG STL-11000 3 CCD | José Francisco Hernández Quico |
| Tecnological University of Pereira<br>Lat +04° 47'<br>Lon 75° 41'<br>Alt 1533 m | W63 | Pereira | Colombia | 40 cm | SBIG ST2000 | Edwin Quintero |

**Table 2.** Observing log of 6478 Gault

| YYYYMMDD | $m_V(1,1,0)$ | t-Tq [d] | Observer | #images |
|---|---|---|---|---|
| 190110 | 14.60±0.04 | -358 | Acosta | 6 |
| 190110 | 14.68±0.04 | -358 | Acosta | 11 |
| 190113 | 14.51±0.03 | -355 | Acosta | 18 |
| 190113 | 14.69±0.03 | -355 | Acosta | 14 |
| 190114 | 14.60±0.03 | -354 | Acosta | 14 |
| 190127 | 14.60±0.03 | -341 | Acosta | 10 |
| 190207 | 14.70±0.02 | -330 | Acosta | 19 |
| 190205 | 14.73±0.02 | -328 | Ferrín | 15 |
| 190211 | 14.76±0.03 | -326 | Acosta | 16 |
| 190213 | 14.69±0.01 | -324 | Acosta | 19 |
| 190213 | 14.76±0.01 | -324 | Acosta | 19 |
| 190215 | 15.00±0.01 | -322 | Fornari | 18 |
| 190216 | 15.00±0.01 | -321 | Fornari | 21 |
| 190217 | 15.00±0.01 | -320 | Fornari | 23 |
| 190218 | 14.87±0.01 | -319 | Fornari | 17 |
| 190222 | 14.87±0.01 | -315 | Fornari | 35 |
| 190225 | 14.81±0.02 | -312 | Acosta | 43 |
| 190225 | 14.88±0.02 | -312 | Fornari | 12 |
| 190227 | 14.92±0.02 | -310 | Fornari | 44 |
| 190315 | 14.84±0.01 | -294 | Acosta | 11 |
| 190316 | 14.88±0.01 | -293 | Acosta | 56 |
| 190325 | 14.77±0.02 | -284 | Fornari | 11 |
| 190326 | 14.97±0.01 | -283 | Fornari | 20 |
| 190327 | 14.81±0.01 | -283 | Acosta | 35 |
| 190329 | 14.84±0.02 | -281 | Acosta | 31 |
| 190331 | 14.75±0.01 | -279 | Hernández | 24 |
| 190406 | 15.18±0.02 | -273 | Acosta | 17 |
| 190407 | 15.13±0.01 | -272 | Acosta | 16 |
| 190409 | 15.44±0.02 | -270 | Fornari | 16 |
| 190409 | 15.89±0.01 | -270 | Acosta | 36 |
| 190409 | 15.21±0.01 | -270 | Hernández | 9 |
| 190427 | 16.15±0.02 | -252 | Acosta | 7 |
| 190430 | 15.90±0.01 | -248 | Fornari | 21 |
| 190501 | 15.67±0.02 | -247 | Fornari | 27 |
| 190508 | 16.20±0.05 | -240 | Canive | 10 |
| 190510 | 15.79±0.02 | -238 | Canive | 11 |
| 190515 | 15.73±0.02 | -233 | Canive | 8 |
| 190516 | 16.01±0.01 | -232 | Fornari | 31 |
| 190525 | 16.11±0.01 | -223 | Fornari | 11 |
| 190528 | 15.89±0.02 | -220 | Quintero | 5 |
| 190608 | 16.10±0.05 | -209 | Fornari | 17 |

**Table 3**. Physical parameters of 6478 determined in this work.

| object | $m_V(1,1,0)$ | D[km] | Asec(q) | Δt (active) [d] | β [mag/°] |
|---|---|---|---|---|---|
| 133P | 15.83±0.05 | 4.2±0.8 | -2.4±0.01 | 150 | 0.044 |
| 62412[2] | 14.18±0.25 | 3.9±0.3 | ----- | ----- | ---- |
| 6478 single | 16.11±0.05 | 2.9 to 3.9 | Z1 -3.4 | ~70 | 0.0 |
| ---- | ---- | ---- | Z2 -2.1 | ~25 | ---- |
| ---- | ---- | ---- | Z3 -1.4 | ~88 | ---- |
| ---- | ---- | ---- | Z4 -1.4 | ~272 | ---- |
| ---- | ---- | ---- | Z5 -1.4 | ~200 | ---- |
| ---- | ---- | ---- | Z6 -1.4 | ~200 | ---- |
| 6478 binary | 16.23±0.05 | 3.7 & 1.3 | ---- | ---- | ---- |
| | | 3.1 & 1.1 | ---- | ---- | ---- |
| | | 2.8 & 1.0 | ---- | ---- | ---- |

This Table gives the absolute magnitude, diameter, amplitude of the SLC, activity interval, and phase coefficient.

1- 6478 has several zones of activity, Z1-Z6. Because Z1 and Z2 have the largest amplitude and the shortest duration they might be the result of an impact.

2- Data from Sheppard and Trujillo (2014).

**Table 4.** Rotational Periods

| Object | Rotational Period [h] | Reference |
|---|---|---|
| 133P | 3.471±0.001 | Hsieh et al. (2014) |
| 6478 | 3.360±0.005 | This Work |
| 62412 | 3.33±0.01 | Sheppard and Trujillo (2014) |

**Table 5.** Orbital Elements and Tisserand Parameter of 133P, 6478 and 62412
Asteroids have $T_J > 3.0$; Comets of the Jupiter family have $2.0 < T_J < 3.0$

| Object | q[AU] | Q[AU] | e | i[°] | a[AU] | Tq yyyymmdd | Porbit [y] | Family Tiss |
|---|---|---|---|---|---|---|---|---|
| 6478 Gault | 1.86 | 2.75 | 0.19 | 22.8 | 2.30 | 2020 01 03 | 3.5 | Phocaea[1] 3.461 |
| 133P | 2.66 | 3.66 | 0.16 | 1.39 | 3.16 | 2018 09 21 | 5.62 | Themis[2] 3.185 |
| 62412 | 2.90 | 3.40 | 0.81 | 4.73 | 3.15 | 2018 10 29 | 5.59 | Hygeia[3] 3.197 |

1- Nesvorny (2015). 2- Hisieh et al. (2004). 3- Sheppard and Trujillo (2014).

**Table 6.** Rotational plot: measurements. Add 2.458.000 to get Julian Dates.

| Time | mag | Time | mag | Time | mag | Time | mag |
|---|---|---|---|---|---|---|---|
| 496.58051 | 3.6396 | 539.49945 | 3.5968 | 559.50057 | 3.5426 | 580.36882 | 3.5621 |
| 496.58436 | 3.6436 | 539.50329 | 3.5851 | 559.50371 | 3.5721 | 580.37132 | 3.5426 |
| 496.58833 | 3.6188 | 539.50713 | 3.5744 | 559.50684 | 3.4772 | 580.37382 | 3.5012 |
| 496.59217 | 3.6538 | 539.51097 | 3.5646 | 559.50999 | 3.5062 | 580.37632 | 3.5426 |
| 496.59603 | 3.6152 | 539.51482 | 3.4938 | 559.51314 | 3.5126 | 580.37881 | 3.5478 |
| 496.59987 | 3.5902 | 539.51866 | 3.5126 | 559.51628 | 3.5162 | 580.38131 | 3.5776 |
| 496.60371 | 3.6111 | 539.52251 | 3.5568 | 559.51941 | 3.5248 | 580.38381 | 3.6028 |
| 496.61308 | 3.6202 | 539.52633 | 3.4782 | 559.52255 | 3.5288 | 580.38631 | 3.6476 |
| 496.62809 | 3.5922 | 539.53016 | 3.4281 | 559.52569 | 3.4816 | 580.38881 | 3.5768 |
| 496.63193 | 3.5991 | 559.36165 | 3.4562 | 559.52882 | 3.4931 | 580.39129 | 3.5876 |
| 496.63576 | 3.5521 | 559.36515 | 3.4278 | 559.53196 | 3.5104 | 580.39378 | 3.5972 |
| 496.63959 | 3.5308 | 559.36831 | 3.3998 | 559.53511 | 3.4336 | 580.39628 | 3.5838 |
| 496.64342 | 3.5758 | 559.37145 | 3.5261 | 559.56333 | 3.3884 | 580.39877 | 3.5772 |
| 496.64726 | 3.5451 | 559.37461 | 3.6042 | 559.56647 | 3.3462 | 580.40127 | 3.6552 |
| 496.65109 | 3.5684 | 559.37801 | 3.6281 | 559.56962 | 3.2142 | 580.40375 | 3.6571 |
| 496.65493 | 3.5746 | 559.38116 | 3.6711 | 559.57277 | 3.1648 | 581.36004 | 3.6694 |
| 496.65876 | 3.5918 | 559.38431 | 3.7236 | 559.57591 | 3.1772 | 581.36249 | 3.6754 |
| 496.66261 | 3.5362 | 559.38746 | 3.6732 | 559.57904 | 3.3032 | 581.36494 | 3.7046 |
| 521.51586 | 3.5611 | 559.39066 | 3.6912 | 572.44511 | 3.3772 | 581.36741 | 3.7042 |
| 521.52041 | 3.5552 | 559.39468 | 3.7852 | 572.44758 | 3.5286 | 581.36985 | 3.6751 |
| 521.52497 | 3.5851 | 559.39783 | 3.7586 | 572.45008 | 3.5466 | 581.37231 | 3.6832 |
| 521.52953 | 3.6104 | 559.40099 | 3.6746 | 572.45256 | 3.5992 | 581.37476 | 3.7244 |
| 521.53425 | 3.6124 | 559.40414 | 3.6541 | 572.45503 | 3.6046 | 581.77211 | 3.7031 |
| 521.53881 | 3.6162 | 559.40728 | 3.5948 | 572.45751 | 3.5818 | 581.79721 | 3.6686 |
| 521.54336 | 3.5994 | 559.41043 | 3.5111 | 572.46111 | 3.5254 | 581.38224 | 3.6482 |
| 521.54791 | 3.5954 | 559.41357 | 3.5121 | 572.46247 | 3.5492 | 581.38469 | 3.6131 |
| 521.55246 | 3.5462 | 559.41672 | 3.5321 | 572.46495 | 3.5386 | 581.38721 | 3.4831 |
| 521.55701 | 3.5088 | 559.41988 | 3.5324 | 572.46743 | 3.4892 | 581.38966 | 3.4418 |
| 521.56174 | 3.4746 | 559.42302 | 3.5532 | 572.47054 | 3.5432 | 581.39211 | 3.4324 |
| 521.56631 | 3.4811 | 559.42635 | 3.5806 | 572.47303 | 3.5844 | 581.39456 | 3.4222 |
| 521.57085 | 3.4782 | 559.42948 | 3.5642 | 572.47551 | 3.6058 | 581.39702 | 3.3964 |
| 521.57541 | 3.4824 | 559.43262 | 3.5882 | 572.47798 | 3.6612 | 582.36147 | 3.4818 |
| 521.57996 | 3.5392 | 559.43577 | 3.6318 | 572.48046 | 3.6854 | 582.36464 | 3.5002 |
| 521.58451 | 3.5704 | 559.43893 | 3.6248 | 572.48294 | 3.6422 | 582.36781 | 3.4828 |
| 521.58923 | 3.5918 | 559.44208 | 3.6624 | 572.48541 | 3.6331 | 582.37098 | 3.4796 |
| 539.43564 | 3.6754 | 559.44522 | 3.7211 | 572.48791 | 3.6468 | 582.37415 | 3.5642 |
| 539.43948 | 3.6604 | 559.44837 | 3.6632 | 572.49038 | 3.6034 | 582.37731 | 3.5596 |
| 539.44332 | 3.6516 | 559.45152 | 3.5821 | 572.49286 | 3.5958 | 582.38049 | 3.4544 |
| 539.44717 | 3.6474 | 559.47861 | 3.5492 | 572.49941 | 3.5806 | 582.38365 | 3.4834 |
| 539.45101 | 3.6128 | 559.48175 | 3.4642 | 572.50436 | 3.5481 | 582.38682 | 3.5174 |
| 539.45485 | 3.5192 | 559.48489 | 3.4056 | 572.50931 | 3.5551 | 582.38999 | 3.4796 |
| 539.45869 | 3.5542 | 559.48802 | 3.4194 | 572.51185 | 3.5568 | 582.39321 | 3.4682 |
| 539.46254 | 3.5608 | 559.49116 | 3.5286 | 572.51433 | 3.5328 | 582.39638 | 3.5906 |
| 539.46644 | 3.5752 | 559.49431 | 3.5296 | 580.36381 | 3.5828 | ------------ | ------- |
| 539.47028 | 3.5756 | 559.49743 | 35676 | 580.36631 | 3.5621 | ------------ | ------- |